\documentclass[twocolumn]{aastex62}

\usepackage{multirow}

\submitjournal{ApJL}

\renewcommand{\vec}{\mathbf}

\newcommand{\affdpa}{\affiliation{Department of Physics \& Astronomy, University of Delaware, Newark, DE, USA}}
\newcommand{\affbartol}{\affiliation{Bartol Research Institute, Newark, DE, USA}}
\newcommand{\affssi}{\affiliation{Space Science Institute, Boulder, CO, USA}}
\newcommand{\affswri}{\affiliation{Southwest Research Institute, San Antonio, TX, USA}}
\newcommand{\affgsfc}{\affiliation{NASA Goddard Space Flight Center, Greenbelt, MD, USA}}
\newcommand{\affucla}{\affiliation{University of California, Los Angeles, CA, USA}}
\newcommand{\affunh}{\affiliation{University of New Hampshire, Durham, NH, USA}}
\newcommand{\affdenali}{\affiliation{Denali Scientific, Fairbanks, Alaska, USA}}

\newcommand{\urlmms}{\url{https://lasp.colorado.edu/mms/sdc/}}

\begin{document}

\title{MMS Observations of Beta-Dependent Constraints on\\Ion Temperature-Anisotropy in Earth's Magnetosheath}

\correspondingauthor{Bennett A.~Maruca}
\email{bmaruca@udel.edu}

\author[0000-0002-2229-5618]{Bennett A.~Maruca}
\affdpa{}

\author[0000-0001-8478-5797]{A.~Chasapis}
\affdpa{}

\author[0000-0002-4655-2316]{S.~P.~Gary}
\affssi{}

\author[0000-0002-6962-0959]{R.~Bandyopadhyay}
\affdpa{}

\author[0000-0002-7174-6948]{R.~Chhiber}
\affdpa{}

\author[0000-0003-0602-8381]{T.~N.~Parashar}
\affdpa{}

\author[0000-0001-7224-6024]{W.~H.~Matthaeus}
\affdpa{}
\affbartol{}

\author[0000-0003-1861-4767]{M.~A.~Shay}
\affdpa{}
\affbartol{}

\author[0000-0003-0452-8403]{J.~L.~Burch}
\affswri{}

\author[0000-0002-3150-1137]{T.~E.~Moore} 
\affgsfc{}

\author[0000-0001-9249-3540]{C.~J.~Pollock}
\affdenali{}

\author[0000-0001-8054-825X]{B.~J.~Giles}
\affgsfc{}

\author{W.~R.~Paterson}
\affgsfc{}

\author{J.~Dorelli}
\affgsfc{}

\author[0000-0003-1304-4769]{D.~J.~Gershman}
\affgsfc{}

\author{R.~B.~Torbert}
\affunh{}

\author[0000-0003-1639-8298]{C.~T.~Russell}
\affucla{}

\author[0000-0001-9839-1828]{R.~J.~Strangeway}
\affucla{}

\date{\today}

\begin{abstract}

Protons (ionized hydrogen) in the solar wind frequently exhibit distinct temperatures ($T_{\perp p}$ and $T_{\parallel p}$) perpendicular and parallel to the plasma's background magnetic-field.  Numerous prior studies of the interplanetary solar-wind have shown that, as plasma beta ($\beta_{\parallel p}$) increases, a narrower range of temperature-anisotropy ($R_p\equiv T_{\perp p}\,/\,T_{\parallel p}$) values is observed.  Conventionally, this effect has been ascribed to the actions of kinetic microinstabilities.  This study is the first to use data from the Magnetospheric Multiscale Mission (MMS) to explore such $\beta_{\parallel p}$-dependent limits on $R_p$ in Earth's magnetosheath.  The distribution of these data across the $(\beta_{\parallel p},R_p)$-plane reveals limits on both $R_p>1$ and $R_p<1$.  Linear Vlasov theory is used to compute contours of constant growth-rate for the ion-cyclotron, mirror, parallel-firehose, and oblique-firehose instabilities.  These instability thresholds closely align with the contours of the data distribution, which suggests a strong association of instabilities with extremes of ion temperature anisotropy in the magnetosheath.  The potential for instabilities to regulate temperature anisotropy is discussed.

\end{abstract}


\section{Introduction} \label{sec:intro}

The solar wind consists of the highly ionized, strongly magnetized plasma that flows supersonically from the Sun's corona out into deep space.  As the plasma approaches Earth, it crosses a bow shock, which reduces it to subsonic speeds and typically causes significant increases in the plasma's density and temperature.  This region of subsonic solar-wind plasma between the bow shock and the Earth's magnetosphere is known as the magnetosheath.

The vast majority of ions in solar-wind plasma are protons, but $\alpha$-particles still typically comprise at least a few percent of ions.  Because of the plasma's high temperature and low density, rates of Coulomb collisions remain low, and ions often deviate from thermal equilibrium.  For example, different ion species frequently exhibit distinct temperatures and bulk velocities.  Furthermore, because of the solar wind's strong magnetic-field, $\vec{B}$, the transport of energy in this plasma is direction dependent, which gives rise to temperature anisotropy.  For protons (ionized hydrogen), which constitute the most abundant ion species, this effect is quantified by the ratio
\begin{equation} \label{eqn:r}
R_p \equiv \frac{T_{\perp p}}{T_{\parallel p}}
\end{equation}
where $T_{\perp p}$ and $T_{\parallel p}$ are respectively the proton temperatures perpendicular and parallel to $\vec{B}$.  A value of $R_p = 1$ corresponds to temperature isotropy, which is a property of the equilibrium state.

As detailed below in Section \ref{sec:back}, numerous studies have shown that the distribution of $R_p$-values observed in the solar wind depends strongly on the parameter
\begin{equation} \label{eqn:beta}
\beta_{\parallel p} \equiv \frac{n_p\,k_B\,T_{\parallel p}}{B^2/\left(2\,\mu_0\right)}
\end{equation}
where $n_p$ is the proton density, $k_B$ the Boltzmann constant, and $\mu_0$ the vacuum permeability.  Essentially, $\beta_{\parallel p}$ is the ratio of the parallel proton pressure to the magnetic pressure.  These studies have revealed that, for progressively larger $\beta_{\parallel p}$-values, the range of $R_p$-values observed in the solar wind narrows.

This study is the first to use data from the Magnetospheric Multiscale Mission (MMS) to study how proton temperature-anisotropy, $R_p$, varies with $\beta_{\parallel p}$ in Earth's magnetosheath and to explore what role instabilities play in the process.  Section \ref{sec:back} details prior observations of correlation of $R_p$ with $\beta_{\parallel p}$ and overviews the theory of kinetic microinstabilities.  Sections \ref{sec:obs} and \ref{sec:anls} respectively describe the specific MMS observations used in this study and how they were analyzed.  A discussion of the results of this analysis is presented in Section \ref{sec:disc}.

\section{Background} \label{sec:back}

As noted in Section \ref{sec:intro}, numerous prior studies (see below) have shown that the values of proton temperature anisotropy, $R_p$, observed in the solar wind depend on the value of $\beta_{\parallel p}$.  Specifically, as the value of $\beta_{\parallel p}$ increases, a progressively narrower range of $R_p$-values is observed.  Conventionally, this effect has been ascribed to the actions of kinetic microinstabilities.  This section provides a brief overview of these instabilities and the evidence of their role in the solar wind.  The textbooks by \citet{book:gary93} and \citet{book:treumann97} as well as the review articles by \citet{schwartz80} and \citet{yoon17} provide a far more detailed treatment of the subject.

A kinetic microinstability is an instability associated with the velocity distribution function (VDF) of one or more particle species in the plasma.  Whenever a VDF deviates from the entropically preferred Maxwellian VDF, free energy is present in the plasma on kinetic scales.  If such a deviation becomes sufficiently large, a fluctuation may develop into a growing mode: its amplitude may increase exponentially until particles are scattered in phase space and the plasma is driven to a state closer to thermal equilibrium.  In this perspective, the thresholds of microinstabilities can act as limits on the plasma's deviation from equilibrium.

Proton temperature anisotropy may drive various instabilities if the value of $R_p \equiv T_{\perp p}\,/\,T_{\parallel p}$ deviates sufficiently from unity.  When $T_{\perp p} > T_{\parallel p}$, the ion-cyclotron instability and/or mirror instability may develop.  Both arise from electromagnetic fluctuations, but, for the former, these modes propagate parallel to the background magnetic-field, $\vec{B}$, and, for the latter, the modes have an oblique orientation to $\vec{B}$ and are non-propagating (i.e., have a phase speed of zero).  The case of $T_{\parallel p} > T_{\perp p}$ may drive the parallel and/or oblique firehose instabilities, which, as their names imply, respectively arise from electromagnetic modes oriented parallel or obliquely to $\vec{B}$.

Various studies \citep[e.g.,][]{book:gary93,hellinger06,maruca12} based on linear Vlasov theory have predicted that the threshold $R_p$-value for each of these four instabilities depends strongly on $\beta_{\parallel p}$ and approaches $R_p=1$ at large $\beta_{\parallel p}$-values.  Consequently, a common method for searching for the effects of these instabilities in a plasma involves plotting the distribution of a large sample of observations over the $(\beta_{\parallel p},R_p)$-plane\footnote{Such a plot is sometimes informally referred to as a ``Brazil plot'' after the shape of the distribution often seen in the solar wind; see, e.g., Figures~\ref{fig:count} and \ref{fig:prob} in this work and Figure~1 by \citet{hellinger06}.}.  The alignment of the data distribution with contours of constant growth-rate, $\gamma$, for a given instability are conventionally interpreted as strong evidence for the action of that instability on the observed plasma.

The $(\beta_{\parallel p},R_p)$-plane has been used extensively to study the impact of anisotropy-driven instabilities on protons in the interplanetary solar-wind.  \citet{kasper02} were among the first to apply this type of analysis to a large sample of solar-wind measurements.  Later, the seminal work of \citet{hellinger06} revealed that the oblique (i.e., mirror and oblique-firehose) instabilities are more active in limiting $R_p$-values than the parallel (i.e., ion-cyclotron and parallel-firehose) instabilities -- even for $\beta_{\parallel p}$-values for which linear Vlasov theory predicts that the latter provide a stricter constraint on $R_p$ than the former.  Subsequent studies supported this conclusion and further developed the idea that instabilities play an important role in how solar-wind plasma expands, develops fluctuations, and is heated \citep{matteini07,bale09,maruca11}.  More recently, the work of \citet{osman12,osman13} and \citet{servidio14} moved beyond linear Vlasov theory to show that $\beta_{\parallel p}$-dependent constraints on $R_p$ are also associated with quantitative measures of turbulence in the solar-wind plasma.  %These observations are consistent with various simulations that have shown an association between turbulent activity and temperature-anisotropy instabilities

Despite this large body of work on the interplanetary solar-wind, few recent efforts have been made to explore proton temperature-anisotropy instabilities in the magnetosheath.  The 1990's did see a series of important studies that used observations from various spacecraft to explore proton temperature-anisotropy in the magnetosheath, but many relied on relatively small datasets and largely focused only on the case of $T_{\perp p} > T_{\parallel p}$.  \citet{anderson94} found evidence of both ion-cyclotron and mirror modes in the magnetosheath but in distinct regions therein.  A subsequent study \citep{anderson96} identified and explored a series of periods of enhanced electromagnetic ion-cyclotron (EMIC) activity and found indications of instability-driven limits on $R_p$.  \citet{gary95} and \citet{tan98} likewise found strong indications that the ion-cyclotron instability acts in the magnetosheath.  \citet{phan94,phan96} focused on the mirror instability and revealed that it too plays a role in regulating the temperature anisotropy of magnetosheath protons.

This study is the first to use MMS observations of the magnetosheath to search for $\beta_{\parallel p}$-dependent limits on $R_p$-values and to explore what role kinetic microinstabilities play in generating them.  Though prior studies \citep[e.g.,][]{wang17} have already identified evidence of unstable modes in MMS data, the analysis herein focuses on understanding how instabilities affect proton temperature.

\section{Observations} \label{sec:obs}

The Magnetospheric Multiscale Mission (MMS) \citep{burch16a} consists of a constellation of four, virtually identical spacecraft that carry instruments optimized for measuring the plasma in and around Earth's magnetosphere (including the magnetosheath) with unprecedented resolution.  All MMS data products are publicly available via the mission's Science Data Center (SDC): \urlmms{}.

For this study, proton data were taken from the Dual Ion Spectrometer (DIS), which is a part of the Fast Plasma Investigation (FPI) \citep{pollock16}.  In burst mode, FPI/DIS returns one distribution of the ion energies every $150\ {\rm ms}$.  Each distribution provides values for the proton moments (including density, $n_p$, and perpendicular and parallel temperature, $T_{\perp p}$ and $T_{\parallel p}$), which are hosted on the SDC.

The magnetic-field data were derived from the Flux Gate Magnetometers (FGM) \citep{russell16} in the FIELDS instrument suite \citep{torbert16}.  In burst mode, FIELDS/FGM provides measurements of the vector magnetic field, $\vec{B}$, at a cadence of $128\ {\rm Hz}$.

\section{Analysis} \label{sec:anls}

\begin{table*}
\caption{\label{tab:prds} Periods of MMS burst-mode data used in this study.}
\begin{center}
\begin{tabular}{lr@{ -- }lrrrrr}
\hline
                         & \multicolumn{2}{c}{}            & \multicolumn{5}{c}{Median Conditions} \\
\multicolumn{1}{c}{Date} & \multicolumn{2}{c}{Time Period} & \multicolumn{1}{c}{$n_p\ [{\rm cm^{-3}}]$} & \multicolumn{1}{c}{$T_{\parallel p}\ [{\rm eV}]$} & \multicolumn{1}{c}{$R_p$} & \multicolumn{1}{c}{$B\ [{\rm nT}]$} & \multicolumn{1}{c}{$\beta_{\parallel p}$} \\
\hline
2016-01-11               & 00:57:04 & 01:00:34             &  52. & 206. & 1.09 & 27.1 &  6.3 \\
2016-01-24               & 23:36:14 & 23:47:34             &  33. & 342. & 0.99 & 18.8 & 12.6 \\
2016-10-25               & 09:45:54 & 09:54:34             & 187. & 282. & 1.02 & 43.5 & 11.3 \\
2017-01-18               & 00:45:53 & 00:49:43             & 198. & 115. & 0.97 & 26.9 & 12.9 \\
2017-01-27               & 08:02:03 & 08:08:03             &  15. & 655. & 1.01 & 20.1 &  9.6 \\
2017-11-23               & 03:57:43 & 04:01:03             &  22. & 241. & 1.05 & 15.6 &  7.9 \\
\hline
\end{tabular}
\end{center}
\end{table*}

\begin{figure}
\begin{center}
\includegraphics[width=3.0in]{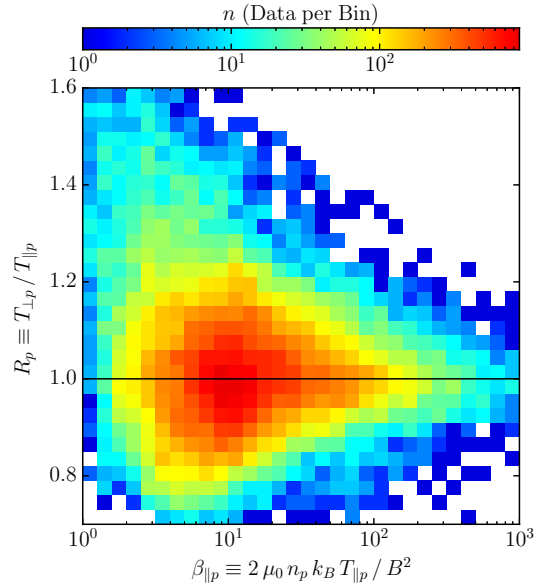}
\caption{\label{fig:count} Plot of the distribution of $(\beta_{\parallel p},R_p)$-values in the MMS dataset specified in Table \ref{tab:prds}.}
\end{center}
\end{figure}

\begin{figure*}
\begin{center}
\begin{tabular}{c@{\hspace{0.25in}}c}
\includegraphics[width=3.0in]{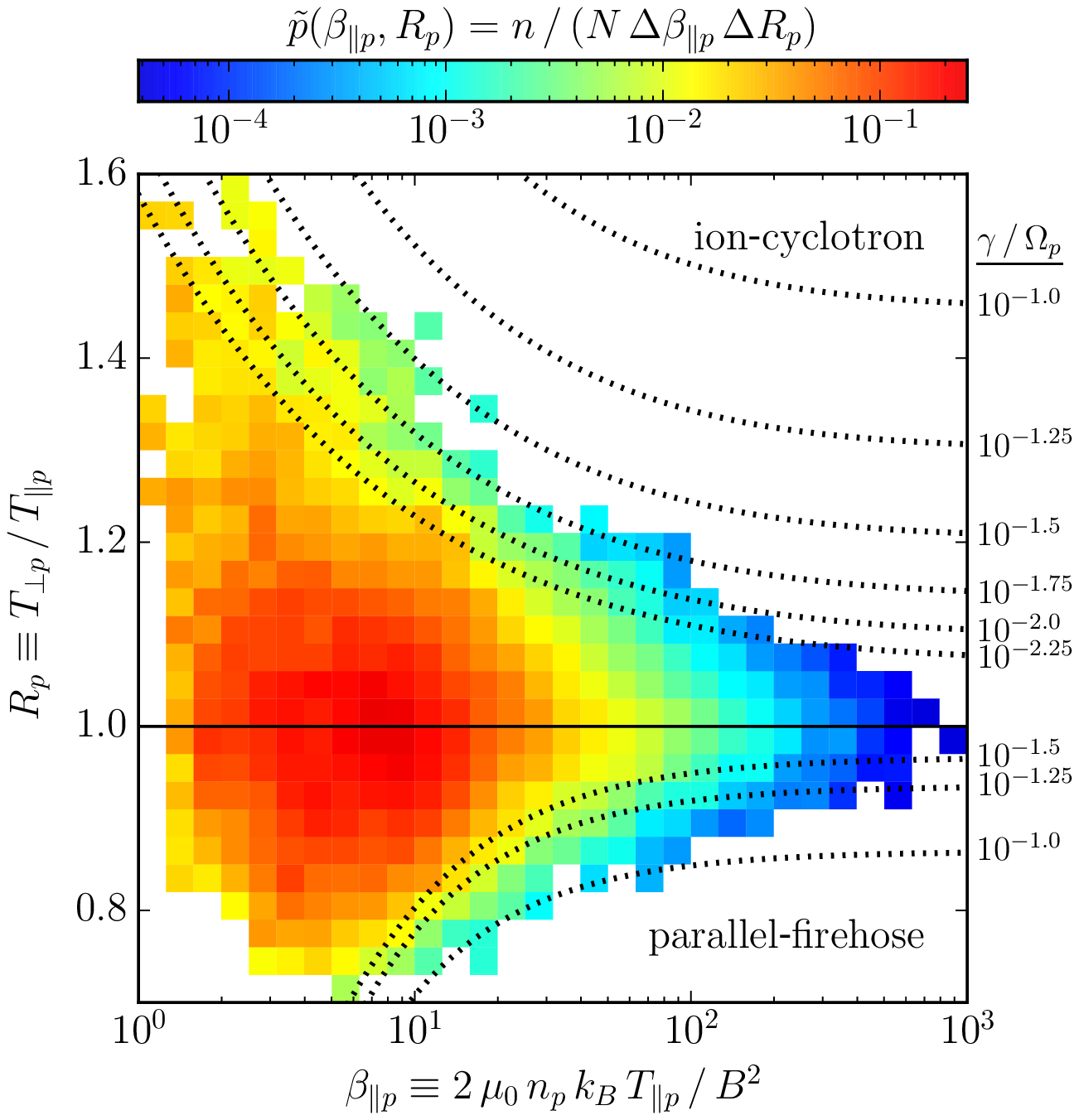} & \includegraphics[width=3.0in]{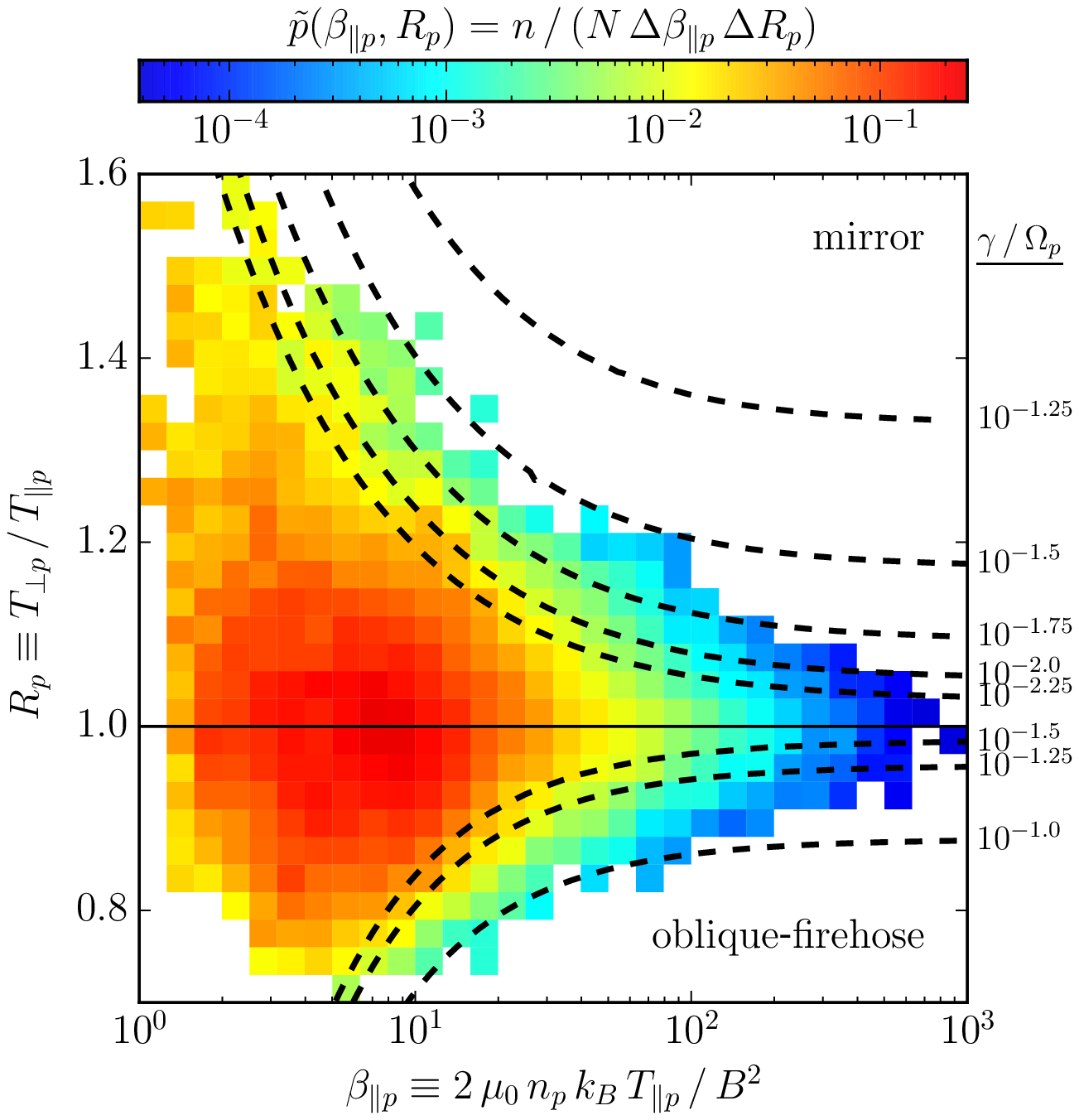} \\
\end{tabular}
\caption{\label{fig:prob} Two plots of the estimated probability density, $\tilde{p}$, of $(\beta_{\parallel p},R_p)$-values (see Equation \ref{eqn:prob}) for the MMS dataset specified in Table \ref{tab:prds}.  These plots are identical except for the overlaid curves, which show contours of constant growth rate for different instabilities.  The dotted curves (left) show the parallel instabilities: the ion-cyclotron ($R_p>1$) and parallel-firehose ($R_p<1$).  The dashed curves (right) show the oblique instabilities: the mirror ($R_p>1$) and oblique-firehose ($R_p<1$).  Each contour is labeled with its growth rate, $\gamma$, in units of the proton cyclotron-frequency, $\Omega_p$ (see Equation \ref{eqn:omega}).}
\end{center}
\end{figure*}

This study utilized a dataset consisting of MMS burst-mode measurements from six distinct periods, which are listed in Table \ref{tab:prds}.  These periods were chosen for their use in previous studies \citep{chasapis17b,chasapis18} and to provide a range of magnetosheath locations and conditions.  Two of these periods had particularly high densities ($n_p > 100\ {\rm cm}^3$), but the ion count-rates remained $\lesssim 4\ {\rm MHz}$ and were localized in the energy scans.  Thus, significant saturation of the instrument is unlikely to have occurred \citep{pollock16,inbo:mcfadden09}, and the DIS estimates of temperature and density were deemed to be of sufficient quality for this study.

No multi-spacecraft techniques were employed in this study: each of the four MMS spacecraft was treated as an independent observer.  Each spacecraft's ion and magnetic-field measurements were synchronized, and the latter were averaged-down to match the $150$-ms cadence of the former.  This ultimately produced a dataset consisting of $N=58,\!510$ measurements of the proton moments and the magnetic-field.

Figure \ref{fig:count} shows the distribution of $(\beta_{\parallel p},R_p)$-values in this dataset.  To generate this plot, the $(\beta_{\parallel p},R_p)$-plane was divided into a grid of bins: $30$ $\beta_{\parallel p}$-bins (logarithmically spaced from $1$ to $10^3$) by $30$ $R_p$-bins (linearly spaced from $0.7$ to $1.6$).  The color of each bin indicates, on a logarithmic scale, $n$, the number of data from the dataset that it contains.

Since the binning of data in Figure \ref{fig:count} was arbitrary, Figure \ref{fig:prob} was generated to show a more fundamental quantity than ``counts per bin.''  Note that the two plots in Figure \ref{fig:prob} are identical except for the overlaid curves, which are addressed below.  The plots themselves were generated by binning the dataset according to the same method used for Figure \ref{fig:count}.  Bins with $n<10$ data were deemed statistically insignificant and suppressed.  Following the method of \citet{maruca11}, the plots show, for the statistically significant bins,
\begin{equation} \label{eqn:prob}
\tilde{p}(\beta_{\parallel p},R_p) = \frac{n}{N\,\Delta\beta_{\parallel p}\,\Delta R_p} \ ,
\end{equation}
where $n$ is the number of data in the bin, $N$ is the total of number of data in the dataset, and $\Delta\beta_{\parallel p}$ and $\Delta R_p$ are the widths of the bin along each axis.  Thus, each value $\tilde{p}(\beta_{\parallel p},R_p)$ estimates the value of $p(\beta_{\parallel p},R_p)$, the probability density of observing the corresponding pair of $(\beta_{\parallel p},R_p)$-values.

The overlaid curves in Figure \ref{fig:prob} show contours of constant instability growth-rate, $\gamma$, across the $(\beta_{\parallel p},R_p)$-plane.  As is typical, $\gamma(\beta_{\parallel p},R_p)$ is taken to be the growth rate of the fastest-growing mode for that set of $(\beta_{\parallel p},R_p)$-values and is normalized to the proton cyclotron-frequency,
\begin{equation} \label{eqn:omega}
\Omega_p = \frac{q_p\,B}{m_p} \ ,
\end{equation}
where $q_p$ and $m_p$ are, respectively, the charge and mass of a proton.  The contours on the left plot correspond to the parallel instabilities: the ion-cyclotron ($R_p>1$) and parallel-firehose ($R_p<1$).  Those on the right plot correspond to the oblique instabilities: the mirror ($R_p>1$) and oblique-firehose ($R_p<1$).  All of these contours were calculated using the same linear-Vlasov software described by \citet{maruca12}, but, for this study, the presence of $\alpha$-particles was neglected.

\section{Discussion} \label{sec:disc}

Figures \ref{fig:count} and \ref{fig:prob} strongly indicate the activity of $\beta_{\parallel p}$-dependent constraints on proton temperature-anisotropy, $R_p$, in the Earth's magnetosheath.  For any given $\beta_{\parallel p}$-value, a distribution of $R_p$-values is observed with a mode near $R_p \approx 1$.  Nevertheless, the width of these $R_p$-distributions narrows considerably with increasing $\beta_{\parallel p}$.  Thus, the magnetosheath likely hosts processes that favor isotropic proton-temperatures (limiting both $R_p>1$ and $R_p<1$) and become more active at higher values of $\beta_{\parallel p}$.

These $\beta_{\parallel p}$-dependent constraints on $R_p$ in the magnetosheath are unlikely to simply be an artifact of conditions in the interplanetary solar-wind, where similar constraints have long been observed \citep[e.g.,][]{kasper02,hellinger06}.  As solar-wind plasma crosses Earth's bow shock, it undergoes significant increases in density and temperature, which likewise increases $\beta_{\parallel p}$.  For example, \citet{maruca11} found the mode of $\beta_{\parallel p}$ in near-Earth interplanetary solar-wind to be $\approx 0.7$, but for the magnetosheath plasma used in this study, the mode of $\beta_{\parallel p}$ is $\approx 8$.  Thus, processes local to the magnetosheath must be primarily responsible for the $R_p$-constraints identified in this study.

Figure \ref{fig:prob} suggests that anisotropy-driven microinstabilities may play a role in limiting $R_p$-values in the magnetosheath.  The plots in this figure show that the vast majority of $(\beta_{\parallel p},R_p)$-values from this study's MMS dataset fall within the limits of marginal stability set by linear Vlasov theory.  Indeed, the observed contours of probability density over the $(\beta_{\parallel p},R_p)$-plane align remarkably well with the predicted contours of constant instability grow-rate.  This study provides some of the clearest indications that instabilities limit temperature anisotropy in high-$\beta_{\parallel p}$ space-plasmas such as the magnetosheath.

The growth-rate contours computed for this study (and shown in Figure \ref{fig:prob}) are remarkably similar for parallel and oblique instabilities (left and right plots, respectively).  In contrast, for the lower-$\beta_{\parallel p}$ conditions of the interplanetary solar-wind, these instabilities show much larger differences: especially for the ion-cyclotron and mirror thresholds ($R_p>1$).  Some studies \citep[e.g.,][]{hellinger06,bale09,maruca11} counter-intuitively found that the distribution of $(\beta_{\parallel p},R_p)$-values observed in interplanetary solar-wind more closely align with the mirror threshold even though the ion-cyclotron threshold typically sets a stronger constraint.  No definitive assessment of the relative activity of the mirror and ion-cyclotron in the magnetosheath is possible in this study because the corresponding thresholds are so similar at high-$\beta_{\parallel p}$.  Qualitatively, Figure \ref{fig:prob} seems to show the mirror-instability contours (with their sharper fall-off) align slightly better with the contours of $\tilde{p}(\beta_{\parallel p},R_p)$ than those of the ion-cyclotron instability.  Nevertheless, the difference is so minor that no clear conclusion can be drawn.

The remarkable results of this preliminary study strongly motivate further investigation to understand how temperature-anisotropy constraints arise in the magnetosheath and how they impact the large-scale evolution of the plasma.  The standard view, reflected in much of our discussion above, is that the occurrence of linear instability near the periphery of the $(\beta_{\parallel p},R_p)$-distribution signifies the role of these instabilities in establishing the limits of the distribution.  The present results are consistent with this view and extend the surrounding questions to the magnetosheath.  Future work might profitably employ a substantially larger dataset of MMS observations to enable more definite conclusions about which instabilities (e.g., the mirror versus the ion-cyclotron) principally constrain $R_p$-values.  Further data could also be used to identify which regions of the magnetosheath host the most active constraints on proton temperature-anisotropy.  Additionally, MMS measurements of other ion species and of electrons should be explored for evidence of $\beta$-dependent constraints on their temperature anisotropies.  Indications of such constraints have already been identified in the magnetosheath for electrons \citep{gary05} and in the interplanetary solar-wind for $\alpha$-particles \citep{maruca12,chen16} and electrons \citep{stverak08,chen16}.

Beyond studies of various regions and particles, efforts should be made to understand how the occurrence of temperature-anisotropy constraints correlate with turbulent structures and fluctuations.  Such correlations have already been observed in the interplanetary solar-wind \citep{osman12,osman13} and identified in turbulence simulations of plasma at intermediate $\beta_{\parallel p}$ \citep{servidio14} and at high $\beta_{\parallel p}$ \citep{kunz14,kunz16}.  Associations of these constraints with regions of enhanced turbulence and coherent structure raise a somewhat different possibility -- that the coherent structures (e.g., current sheets) generate the extremal parameters that create conditions for instability.  In that case, further study would be needed to distinguish the relative impact that instabilities and coherent structures have on the particle distributions.

\section*{Acknowledgments} \label{sec:ack}

Effort on this study at the University of Delaware was partially supported by NASA HSR Grant NNX17AI25G, NASA LWS Grant NNX17AB79G, and MMS through NASA Grant NNX14AC39G.  W.H.M.~is a member of the MMS Theory and Modeling Team.

This study used Level 2 FPI and FIELDS data products in cooperation with the instrument teams and in accordance with their guidelines.  These and all MMS data are available at \urlmms{}.  We thank the SDC, FPI, and FIELDS teams for their assistance with this study.

The preparation of this manuscript made use of the SAO/NASA Astrophysics Data System (ADS).

\end{document}